\documentclass[letterpaper,prl,twocolumn,showpacs,superscriptaddress]{revtex4}

\usepackage{graphicx,bm}

\begin{document}

\title{Room temperature ferromagnetism in Cr-doped hydrogenated amorphous Si films}

\author{Jia-Hsien Yao}
\affiliation{Department of Materials Science and Engineering,
National Tsing Hua University, Hsinchu 30013, Taiwan, ROC}

\author{Hsiu-Hau Lin}
\affiliation{Department of Physics, National Tsing-Hua University,
Hsinchu 30013, Taiwan, ROC}
\affiliation{Physics Division, National Center for Theoretical Sciences,
Hsinchu 30013, Taiwan, ROC}

\author{Tsung-Shune Chin}
\email{tschin@mx.nthu.edu.tw}
\affiliation{Department of Materials Science and Engineering, Feng Chia University, Taichung, 40724, Taiwan, ROC}
\affiliation{National Nano-Devices Lab., National Applied Research Laboratories, 300, Hsinchu, Taiwan, ROC}
\date{\today}

\begin{abstract}
Ferromagnetism above room temperature was observed in Cr-doped hydrogenated amorphous silicon films deposited by rf magnetron-sputtering.  Structure analysis reveals that films are amorphous without any detectable precipitates up to the solubility limit 16 at\% Cr.  Experimental results suggest that hydrogenation has a dramatic influence on magnetic properties, electrical conductivity and carrier concentration in the thin films. Pronounced anomalous Hall effect and magnetization curve both suggest the origin of the ferromagnetism may arises from percolation of magnetic polarons.
\end{abstract}
\maketitle

Diluted magnetic semiconductor (DMS) is among the best candidates to realize spintronics devices with the potentials to manipulate both charge and spin degrees of freedom in one material.\cite{Ohno01} One of the crucial criteria for its practical applications is to maintain robust ferromagnetism near/above room temperature. Even through fabrication of DMS has been achieved by doping transition metals into III-V semiconductors, such as Ga$_{1-x}$Mn$_x$As and In$_{1-x}$Mn$_x$As, the Curie temperatures remain low. On the other hand, to enjoy the smooth integration with existing industry of modern electronics, doping crystalline-Si (c-Si) and crystalline-Ge (c-Ge) seems to kick off another route. For instance, Park \textit{et al.}\cite{Park02} reported that the Curie temperature of Mn$_x$Ge$_{1-x}$ deposited by molecule beam epitaxy increases linearly with the Mn concentration and goes up to 116 K at $x= 0.35$. However, the reported Curie temperatures in c-Si and c-Ge based DMSs are rather scattered, ranging from 116 K to 400 K in current literature\cite{Park02,Chen07,YXChen07,Jamet06,Gareev06}.

The challenge of turning Si/Ge crystalline hosts into DMS lies in the extremely low solubility of the magnetic dopants. This difficulty can be removed by using amorphous silicon (a-Si) that has a higher solubility. However, Liu \textit{et al.}\cite{Liu06} investigated Si$_{1-x}$Mn$_x$:B thin films deposited by co-sputtering and concluded that the as-deposited amorphous films are not ferromagnetic.  Yu \textit{et al.}\cite{Yu06} reported that amorphous Si$_{1-x}$Mn$_x$ ($x< 0.09$) thin films grown by thermal deposition are also paramagnetic. We suspected that the null results for ferromagnetism may originate from the dangling bonds that trap dopants/carriers and deteriorate the conductivity and other desirable properties required for useful semiconductors.  

\begin{figure}
\centering
\includegraphics[width=8cm]{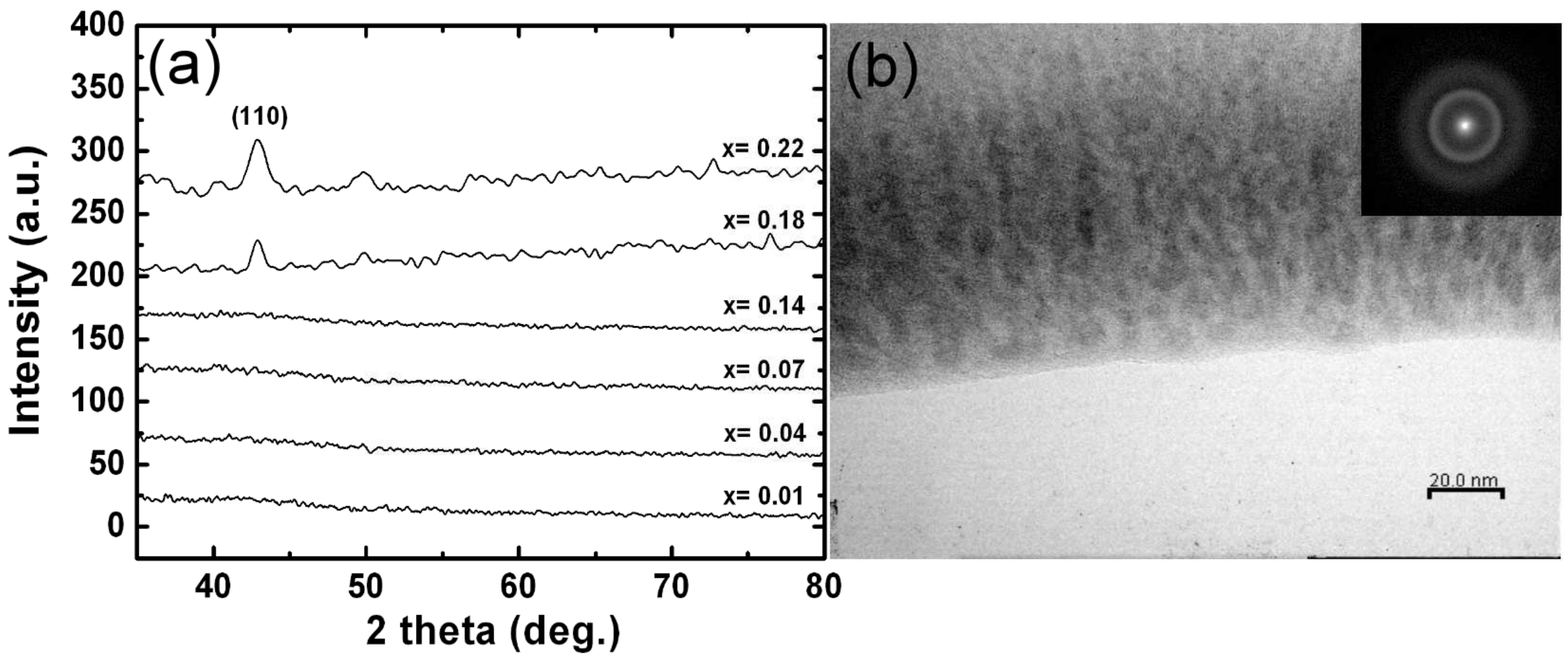}
\caption{\label{Fig1}
(a) XRD patterns from the a-Cr$_x$Si$_{1-x}$:H samples with $x$= 0.01, 0.04, 0.07, 0.14, 0.18, 0.22. (b) High-resolution TEM image of an a-Cr$_{0.14}$Si$_{0.86}$:H sample. The inset shows the corresponding selected-area electron diffraction pattern.}
\end{figure}

The problem associate with dangling bonds can be cured by hydrogenation\cite{Chittick69}, which has been widely used in many applications, such as thin film transistors in liquid crystal display and the p-n junction in solar cells.  In the Letter, we investigate transition metal doped a-Si with or without hydrogenation. Remarkably, hydrogenated samples show robust ferromagnetism even above room temperature while the magnetic order is greatly suppressed without hydrogenation.

Cr-doped hydrogenated amorphous silicon (a-Cr$_x$Si$_{1-x}$:H) films were deposited by rf magnetron co-sputtering onto a glass-substrate using Si and Cr targets simultaneously. All films were deposited in a sputtering atmosphere of Ar/H$_2$ = 80/20 or pure Ar at room temperature.  After deposition, we capped an Al layer on top of the films to prevent oxidation. The thickness of samples was fixed at 1 $\mu$m.  The composition of a-Cr$_x$Si$_{1-x}$ films was varied between $x= 0 \sim 0.22$. Structure of the films was characterized using X-ray diffraction (XRD) and high-resolution transmission electron microscopy (HRTEM) while magnetic measurements were performed by the superconducting quantum interference device (SQUID).  Furthermore, the composition of samples was analyzed by a field-emission electron-probe micro-analyzer (FE-EPMA).  Finally, we also measured the Hall resistivity and the carrier concentration with the Van der Pauw configuration by a Physical Property Measurement System (PPMS).

\begin{table}
\begin{ruledtabular}
\begin{tabular}{ccc}
sample & carrier concentration  & conductivity \\
\hline
Cr$_{0.14}$Si$_{0.86}$:H   & $1.073 \times 10^{22}$ cm$^{-3}$& $2.48 \times 10^{2}$ (ohm-cm)$^{-1}$ \\
Cr$_{0.14}$Si$_{0.86}$ & $1.987 \times 10^{21}$ cm$^{-3}$& $1.54 \times 10^{2}$ (ohm-cm)$^{-1}$\\
\end{tabular}
\end{ruledtabular}
\caption{
Carrier concentration and conductivity of a-Cr$_{0.14}$Si$_{0.86}$ films with or without hydrogen 
\label{tab1}
}
\end{table}

Figure 1(a) shows XRD patterns of a-Cr$_x$Si$_{1-x}$:H films at different doping. At $x=0.18$, Cr (110) peak starts to appear, signaling the clustering above the solubility limit. Since the peak is absent at $x=0.14$, we assume an interpolated solubility limit of roughly 16 at\%. Figure 1(b) shows a typical HRTEM image of an a-Cr$_{0.14}$Si$_{0.86}$:H film.  Although darker and brighter stripes are visible due to localized strain, no precipitate is observed.  The inset diffraction pattern confirms the amorphous nature of the film. However, the magnetic moment associated with each Cr atom, 0.01 $\mu$B/Cr, is unexpectedly small. Since the film is protected by capped Al layer and magnetic Cr oxides are hard to form during sputter deposition, we suspect the small moment arises from formation of clusters beyond detection limit. The existence of such Cr clusters, being antiferromagnetic in nature, quench most Cr spins except the uncompensated moments from surface of the cluster\cite{Neel62}, effectively leading to a rather small moment per Cr atom. But, more experiments are necessary to pin down the exact cause.

\begin{figure}
\centering
\includegraphics[width=7cm]{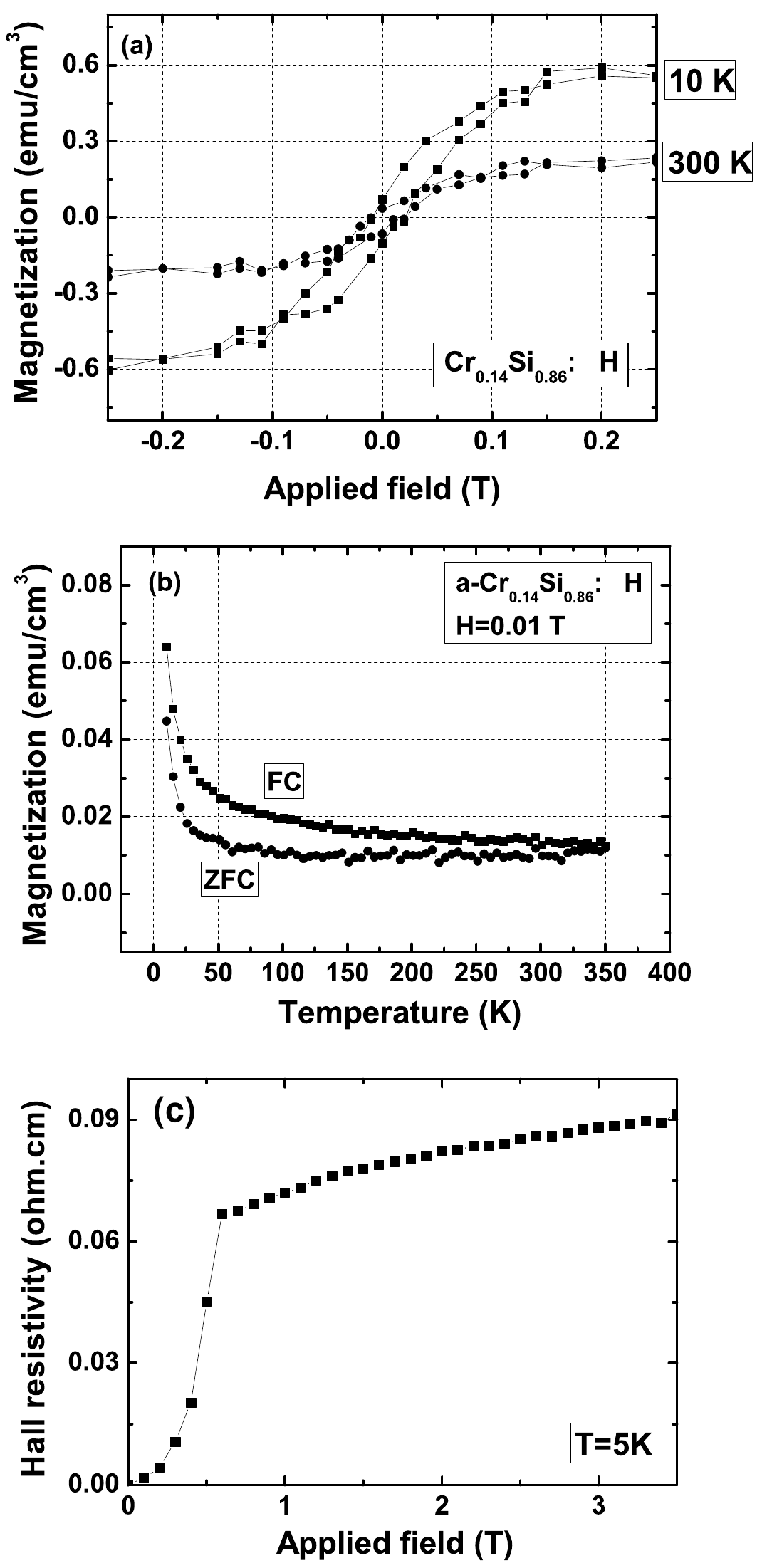}
\caption{\label{Fig2}
(a) M-H curves of an a-Cr$_{0.14}$Si$_{0.86}$:H sample measured at 10 K and 300 K respectively. (b) Magnetization versus temperature in the presence/absence of a magnetic field 100 Oe. (c) Anomalous Hall resistivity measured at 5 K.}
\end{figure}

\begin{figure}
\centering
\includegraphics[width=7cm]{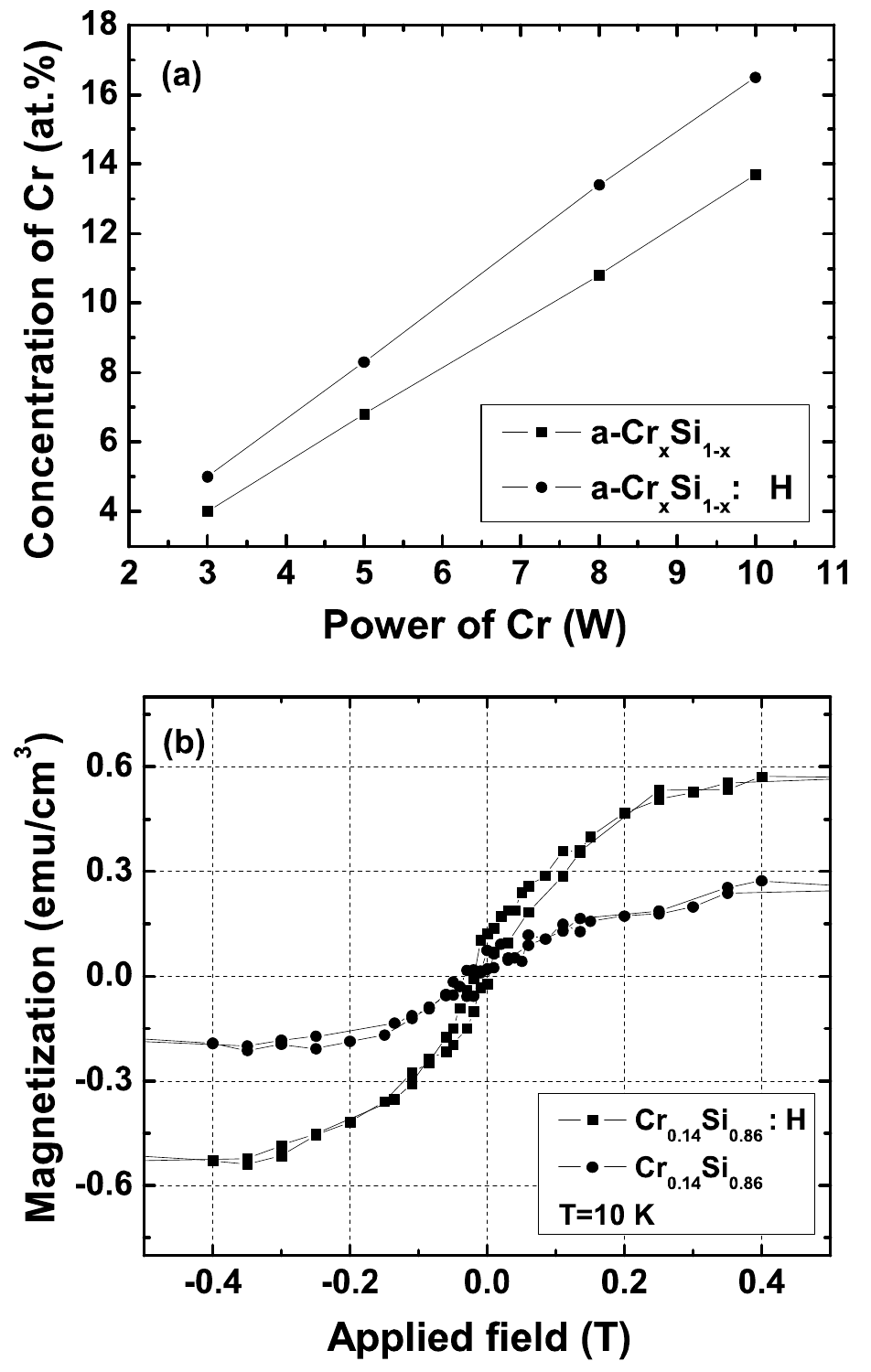}
\caption{\label{Fig3}
(a) The effect of hydrogenation on Cr concentration of a-Si films; (b) M-H curves measured at 10 K of a-Si films of the same Cr concentration with and without hydrogen.}
\end{figure}
\begin{figure}
\centering
\includegraphics[width=7cm]{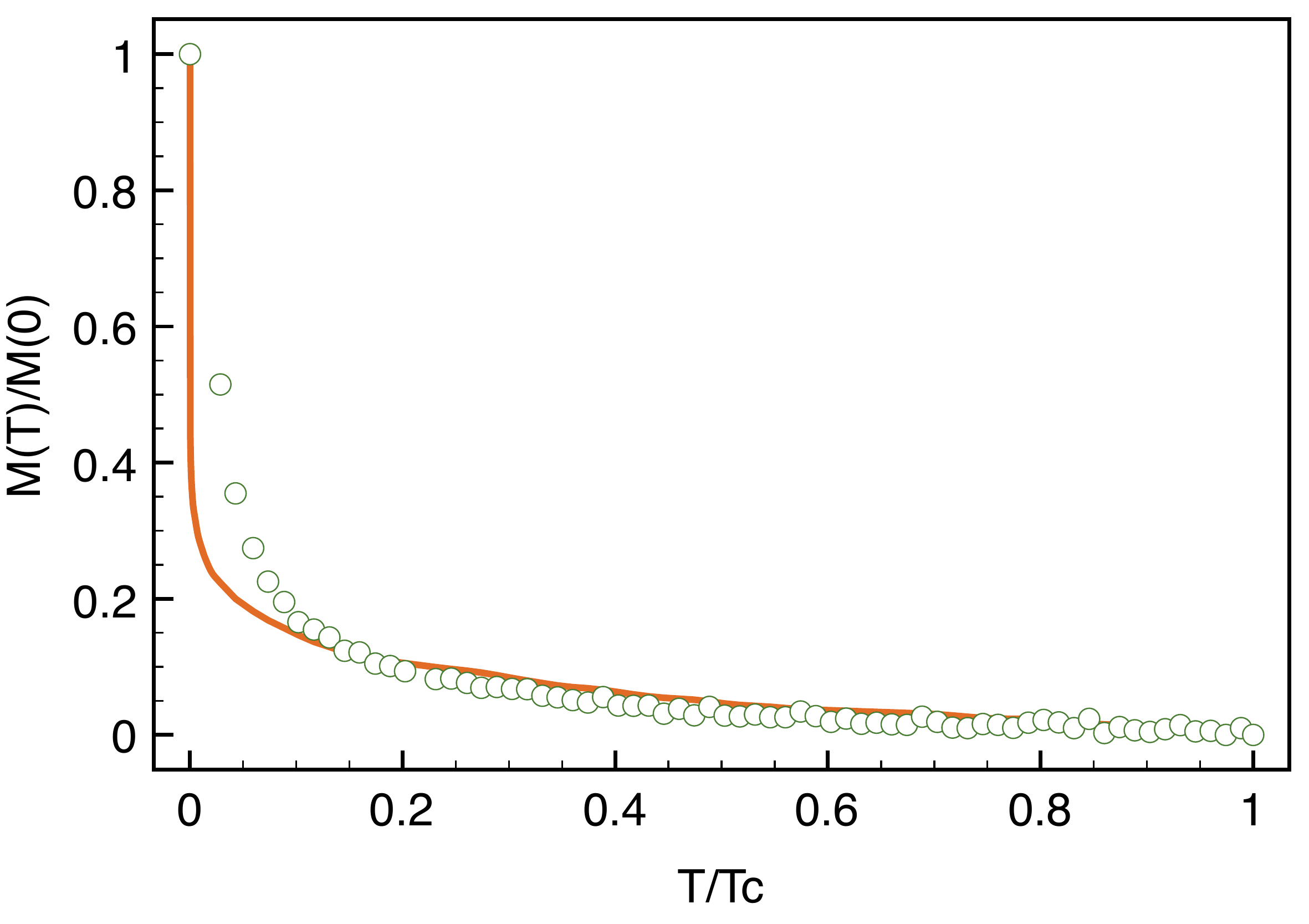}
\caption{\label{Fig4}
Magnetization curve for a-Cr$_{0.14}$Si$_{0.86}$:H sample (green dots) and the theoretical prediction (orange line) from magnetic polaron percolation with $a_B^3 n_h = 2.7 \times 10^{-5}$.}
\end{figure}

We focus on the a-Cr$_{0.14}$Si$_{0.86}$:H sample with optimal magnetic properties. The ferromagnetism is clearly seen in Fig. 2(a) at both 10 K and 300 K. The saturation magnetization is 0.23 emu/cm$^3$ and the coecivity is about 100 Oe at room temperature. Both the field-cooled (FC) and the zero-field-cooled (ZFC) magnetization were measured from 350 K down to 10 K. The crossing point of FC and ZFC curves, serving as a rough estimate for Curie temperature, is clearly above room temperature. In addition, we also measured the Hall resistivity by PPMS in the presence of perpendicular magnetic field up to 3.5 Tesla at T = 5 K.  As shown in Figure 2(c), the anomalous Hall effect (AHE) is manifest, hinting the ferromagnetic order is mediated by charge carriers.

Note that the ferromagnetism sensitively depends on the magnetic doping. We have investigated samples with different doping levels $x=0.11 \sim 0.22$ and found the magnetic order gets strengthened up to the optimal value $x=0.14$ and then suppressed by further doping due to the formation of antiferromagnetic clusters. This is consistent with the results reported by Peak \textit{et al.}\cite{Peak04} for a-Ge$_{1-x}$Cr$_x$ samples. 

Now we turn to the role of hydrogen passivation. Note that, at the same sputtering power, Cr density is higher without hydrogenation as shown in Fig. 3(a). The reduction of sputter yield of Cr is simply due to the presence of hydrogen in the working gases. For comparison, samples at the same Cr doping with and without hydrogenation are listed in Table 1. One notices that carrier concentration of the hydrogenated sample is one order larger than that of the un-hydrogenated, while the conductivity is only a factor of two larger. This implies that hydrogenation increases lots of charge carriers with low mobility. The increase of carrier density coincides with stronger magnetic order in the hydrogenated sample as shown in Fig. 3(b). Hydrogenation enhances saturation magnetization up to 100\%. In fact, aside from the optimal doping $x=0.14$, all other un-hydrogenated a-Cr$_x$Si$_{1-x}$ films show diamagnetic behavior down to 10 K. Our findings explain the absence of ferromagnetism in transition-metal doped a-Si reported earlier\cite{Liu06,Yu06} and establish the importance of hydrogenation to eliminate the dangling bonds in amorphous configuration.

Finally, we would like to explore the origin of the ferromagnetism. First, we assume the holes are itinerant and estimate the Curie temperature by the Zener model within self-consistent Green's function approach.\cite{Sun04,Bunder06} However, with a reasonable exchange coupling between the itinerant and Cr spins ($J \sim 100$ meV nm$^3$), the maximum Curie temperature never exceeds 100 K, far below that observed in our samples. We then turn to the percolation theory of magnetic polaron\cite{Kaminski02} with only one fitting parameter $a_B^3 n_h$, where $a_B$ is the decay length of the magnetic polaron and $n_h$ is the hole density. It is rather remarkable that the magnetization curve at optimal doping $x=0.14$ agrees with the theoretical prediction rather well as shown in Fig. 4. This scenario also agrees with previous observations of the anomalous Hall effect and the presence of low-mobility carriers in the samples.

In conclusion, the Cr-doped hydrogenated a-Si has been found to demonstrate ferromagnetic order at room temperature. It is worth emphasizing that the ferromagnetism is either absent or strongly suppressed without hydrogen passivation.  Only when dangling bonds in the sample are largely hydrogenated, will robust ferromagnetic order be observed.

\end{document}